\documentclass[aip,preprint]{revtex4-1} 
\usepackage{amsmath,amssymb}

\renewcommand{\vec}{\mathbf}
\newcommand{\bea}{\begin{eqnarray}}
\newcommand{\eea}{\end{eqnarray}}
\begin{document}

\title{STATISTICAL DESCRIPTION OF HYDRODYNAMIC PROCESSES IN IONIC MELTS
WITH TAKING INTO ACCOUNT POLARIZATION EFFECTS}

\author{B.~Markiv}
\affiliation{Institute for Condensed Matter Physics NAS of
Ukraine,\\ 1~Svientsitskii Str., 79011 Lviv, Ukraine}
\author{A.~Vasylenko}
\affiliation{Institute for Condensed Matter Physics NAS of
Ukraine,\\ 1~Svientsitskii Str., 79011 Lviv, Ukraine}
\author{M.~Tokarchuk}
\affiliation{Institute for Condensed Matter Physics NAS of
Ukraine,\\ 1~Svientsitskii Str., 79011 Lviv, Ukraine} \affiliation{Lviv Polytechnic National University,  \\12~Bandera Str., 79013 Lviv, Ukraine}
\date{\today}

\begin{abstract}
Statistical description of hydrodynamic processes for molten salts
is proposed with taking into account polarization effects caused
by the deformation of external ionic shells. This description is
carried out by means of the Zubarev nonequilibrium statistical
operator method, appropriate for investigations of both strong and
weak nonequilibrium processes.  The nonequilibrium statistical
operator and the generalized hydrodynamic equations that take into
account polarization processes are received for ionic-polarization
model of ionic molten salts when the nonequilibrium averaged
values of densities of ions number, their momentum, dipole
momentum and total energy are chosen for the reduced description
parameters. A spectrum of collective excitations is investigated
within the viscoelastic approximation for ion-polarization model
of ionic melts.
\end{abstract}



\maketitle

\section{Introduction}

Study of equilibrium and nonequilibrium properties of ionic melts
remains actual from the viewpoint of experimental\cite{1,2,3,4,5,6,7,8,9,10}, theoretical\cite{11,12,13,14,15,16,17,18,19,20,21,22,23,24,26} as well as
computer simulation\cite{27,28,29,29a,29b,30,31,32,33,34,35,21,23,24,26}
investigations. They are very important because of a wide range of
applications in chemical, metallurgical and nuclear technologies\cite{37a,37b,37c,37d,37e,37k}.  It is important to notice the
papers\cite{1,2} where the binary distribution functions for a
number of alkaline-chloride melts and some valence asymmetric ones
were experimentally obtained for the first time by the method of
neutron scattering in the systems with isotope replacement,
developed by I.E.~Enderby and colleagues. Dynamic structure
factors of some ionic melts were received by means of an inelastic
neutron scattering\cite{3,4,5}. Moreover, diffusion,
electroconductivity and viscosity coefficients were examined
experimentally\cite{9} as well. The recent X-rays inelastic
scattering experiments on molten NaCl\cite{6}, NaI\cite{7} and
CsCl\cite{8} have initiated study of collective excitations and
dispersion laws of ionic melts. Theoretical investigations
intended to explain observable phenomena in such systems were
carried out on the basis of the kinetic equations for
one-component plasma\cite{36,37}, the mode coupling theory\cite{38,39,13}, the generalized\cite{12} and the extended\cite{18,19} hydrodynamics, the Zubarev nonequilibrium statistical operator (NSO) method\cite{Zub1,Zub2}, the generalized collective
modes (GCM) approach\cite{40,21,23,24,26} and others. It is
important to note that in consequence of collective excitations
analysis for two-component plasma, five hydrodynamic modes (one
heat, two sound and two mass and charge diffusion modes) and five
relaxation modes were indicated based on the renormalized kinetics
theory\cite{37,39}. Six of those modes describe properties of a
system as neutral and four modes as a charged one. Correlation
functions and corresponding response functions of mass and charge
densities, temperature and momentum divergency fluctuations were
considered\cite{12} using linear hydrodynamics equations, which
takes into account thermoelectric and electrostriction effects.
Statistical hydrodynamics of ionic systems were constructed\cite{18} with help of the NSO method\cite{Zub1,Zub2} based on the extended set of the reduced description parameters, including microscopic densities of particles number, their momentum, total
energy and densities of generalized viscous stress tensor and
energy flow. The obtained equations of extended hydrodynamics are
valid for both strong and weak nonequilibrium processes. Moreover,
this approach was reformulated\cite{19} on the basis of the
generalized Fokker-Planck equation for the collective variables
functional, considering nonlinear hydrodynamic fluctuations in
ionic systems. The approach presented by Zubarev and Tokarchuk\cite{18} permitted to discover mutual influence of heat-viscosity
processes in the time correlation functions ``mass-mass'',
``mass-charge'', ``charge-charge'' and their flows for weakly
nonequilibrium processes in NaCl ionic melt. Theoretical study of
dynamic structure factors, time correlation functions of
longitudinal and transverse currents of mass and charge densities
as well as transport coefficients\cite{11,12,13,14,15,16,17,18}
obtained a qualitative agreement with the results of molecular
dynamics (MD) calculations\cite{27,28,29,29a,29b,30} actively
carried out in the mid 70s of the last century, starting from the
paper by Hansen and McDonald\cite{27}. In that time, the
important results in investigations of structure and dynamical
properties of ionic systems by means of the MD simulations were
obtained. In particular, the equilibrium binary distribution
functions of ionic melts, calculated with the molecular dynamics
method\cite{30} have a good agreement with the experimental data\cite{1,2}. By means of the MD simulations the dynamic structure
factors for the model ionic melts of NaCl\cite{27,28} and RbBr\cite{29}, time correlation functions of ionic velocities,
diffusion and electroconductivity coefficients for ionic melts of
NaCl, LiI, RbCl and others were received\cite{29,29b}. An
interesting fact is that the spectrum of charge density
fluctuations of ionic melts in the long-wavelength limit has a
characteristic peak like a spectrum of longitudinal optic phonons
of ionic crystals. This was shown using the molecular dynamics
simulations\cite{27} and afterward in the theoretical studies\cite{11,12,13,14}. Moreover, it was confirmed by  \textsl{ab initio} (AI) MD calculations in conjunction with the GCM approach\cite{24} and in real experiment\cite{10}.

In the earlier theoretical investigations and MD simulations of the equilibrium and nonequilibrium properties of ionic melts an effective ionic
Mayer-Huggins and Tosi-Fumi potentials\cite{20}
which do not take into account polarization effects  were used.
However, in the real melts outer electrons shells can be polarized
and, therefore, must contribute to an effective interaction. Nowadays, experimental investigations\cite{6,7,8}, MD simulations\cite{30,31,32,34,40b} and AI MD simulations\cite{21,23,24,26} reveal an importance of taking into account polarization processes caused by electronic structure of ions. Polarizability of ions depends on their size and configuration of outer electron shells. The more electrons on the external shell present, the higher ion polarizability is.  In particular, the theoretical study of collective modes dispersion for molten NaCl\cite{21} within the GCM approach based on the both rigid-ion and AI MD simulations showed that, in the domain of small wavenumbers, the rigid-ion model yields higher values of optic modes frequencies compared to AI calculations that takes into account polarization processes. This is the evidence that polarization effects in NaCl (in this case caused mainly by Cl$^-$ ions due to their higher polarizability) can not be neglected. Similar is the situation with molten NaI\cite{23,26,31,Bitrian3} in which the polarizability of I$^-$ ions containing eight electrons in the outer shell is much more then the Na$^+$ ions' one (however, the models with both species of ions to be polarized were considered\cite{32} as well). Similar models for silver halides, in which only halogen ions as well as both cations and anions are assumed to be polarizable, were studied\cite{Bitrian3,Alcaraz}. The importance of polarization effects can be estimated by the value of the high-frequency dielectric permittivity $\varepsilon$ of molten salts. If its value is about 2 or higher ($\varepsilon=1$ corresponds to the rigid-ion model)~--- the polarization effects are essential\cite{26}.

Today polarization effects are described within the three models:
polarization point dipole models (PPDM)\cite{31,Wilson2,Trullas,Bitrian,Bitrian2}, shell models (SM)\cite{41,34,30,29a} and fluctuating charge models (FCM)\cite{22,Wilson1,Ribeiro}. In PPDM an inducted point dipole is added to atomic or ionic positions, whereas in other two models dipole has a finite length. In SM dipole is presented by a pair of
point charges, namely with a positive nucleus and a negative
shell, linked by a harmonic bond; whereas in FCM point
charges are fixed in certain positions and their values can
fluctuate. In particular, influence of polarization effects on
equilibrium properties of ionic melts AgI were considered\cite{40a}. In Ref.\cite{31} polarization effects are described within PPDM model when studying dynamical properties by means of MD simulations. In order to take into account electron shells
deformations a shell model was proposed\cite{23,24} in which
electron shell is presented as an external electron cloud that
interacts with nucleus through the harmonic potential and
repulsive potential at small distance\cite{36}. Based on this
model, the Car-Parrinello AI simulations in which the dynamics of
ionic subsystem was reduced to the pseudo-dynamics of electrons
wave functions within the density functional formalism were
carried out. This permits observation of polarization effects
caused by deformations of outer electron shells at the
\textsl{ab initio} level.

In the present paper a statistical description of hydrodynamic
processes in ionic melts with taking into account polarization
effects caused by deformations of outer electrons shells of
ions is proposed. It is implemented by means of the Zubarev NSO
method\cite{Zub1,Zub2} that permits to study both weak and strong
nonequilibrium processes. In the second section the nonequilibrium
statistical operator and the generalized hydrodynamic equations
taking into account polarization effects are obtained for the
ion-polarization model of ionic melts, when the nonequilibrium
averaged values of densities of ions number $\hat n^a(\vec r)$,
their momentum $\hat{\vec p}^a(\vec r)$, angular momentum
$\hat{\vec s}^a(\vec r)$, total energy $\hat \varepsilon(\vec r)$
and dipole moment $\vec d^a(\vec r)$ are chosen for the reduced
description parameters. The generalized molecular hydrodynamics
equations for ionic melts in the case of weak nonequilibrium are
obtained in the third section. Section~4 contains analytic
expressions for the collective excitations spectrum in the limit
$\vec k\rightarrow 0$, $\omega\rightarrow 0$ on the basis of the
viscoelastic model for ionic melts with taking into account
polarization processes.

\section{Hamiltonian of the system. Nonequilibrium statistical operator}

Let us consider an ion-polarization model of ionic melts which
classically describes ionic subsystem with taking into account
polarization effects. We assume that both positively and
negatively charged ions can be polarized, though in case of NaI molten salt, negatively charged iodine ions only are polarized. In
polarization processes related with electron transition between
orbitals of an atom that turns to a positive ion and a negative
atom of melt that turns to a negative ion, dipole moments of ions
are induced. In consequence of interactions like ``induced dipole
- induced dipole'', ``induced dipole - ion'' and motion dynamics
it leads to induced momenta of dipoles that in turn cause their
rotational motion. Therefore, ions with induced dipole moments
besides translational degrees of freedom possess rotational ones
due to polarization processes. All these degrees of freedom of
ions in melts must be accounted in the full Hamiltonian of the
system. The latter can be presented in the following form:
\bea\label{2}
 H=\sum_a\sum_{j=1}^{N_a}\left(\frac{p_j^2}{2m_a}
 +\frac{1}{2}{\vec
 w}_j^\mathsf{T}{\stackrel{\leftrightarrow}{\mathbf{J}}}_j
{\vec w}_j\right) +U_\mathrm{ion} \,,  \eea
which includes kinetic (translational and rotational energies) and
potential parts; $\vec p_j$ denotes a momentum vector, $m_a$ is a
mass of ion of species $a$, $\vec w_j$ stands for an angular
velocity ($\vec w_j^\mathsf{T}$ is a transposed vector) and
${\stackrel{\leftrightarrow}{\mathbf{J}}}_j$ means an inertia
tensor of the $j$-th  polarized ion (ionic dipole) determined
relatively its center of mass. $U_\mathrm{ion}$ we introduce as
follows\cite{31}:
\bea\label{3}
U_\mathrm{ion}&=&\frac{1}{2}\sum_{a,b}\sum_{i\neq j=1}^{N_a,\, N_b}\Phi_{ab}(r_{ij})-\sum_a\sum_{j=1}^{N_a}\vec
d_j\cdot\vec E_j^q - \frac{1}{2}\sum_a\sum_{j=1}^{N_a}\vec
d_j\cdot\vec E_j^d\nonumber
\\ &&+ \sum_a\sum_{j=1}^{N_a}\frac{d_j^2}{2\alpha_j} +
\sum_{a,b}\sum_{i\neq j=1}^{N_a,\, N_b}f_{ab}(r_{ij})
\frac{Z_{b}e}{r_{ij}^3}\vec r_{ij}\cdot\vec d_i\,. \eea
$\Phi_{ab}(r_{ij})$ is an ion-ion interaction potential:
\bea\label{4} \Phi_{ab}(r_{ij}) = \frac{Z_{a}Z_{b} e^2} {r_{ij}}
+\varphi^{sh.r.}_{ab}(r_{ij}), \eea
where $\varphi^{sh.r.}_{ab}(r_{ij})$ is a short
range potential containing an overlap repulsive part. In MD simulations it was widely used
in the Born-Mayer-Huggins form\cite{41} for alkali halides or in
the form proposed by Parrinello et al.\cite{Parrinello} for
silver halides. $Z_{a}$, $Z_b$ are the valences of the ions of
a corresponding species, $e$ denotes an electron charge. $\vec
d_j$ means a dipole momentum of an ion with a distorted outer
shell
\bea\label{5} \vec d_j = \alpha_j\vec E_j -
\alpha_j\sum_b\sum_{i\neq j=1}^{N_b} f_{ab}(r_{ij}) \frac{Z_{b} e}
{r_{ij}^3}\vec r_{ij}\, , \eea
$\alpha_j$ is a polarizability of the $j$-th ion  in an
electric field
\bea\label{6} && \vec E_j = \vec E_j^d + \vec E_j^q, \\\nonumber
&&\vec E_j^q = \sum_b\sum_{i\neq j=1}^{N_b}\frac{Z_{b} e} {r_{ij}^3}\vec
r_{ij}, \\\nonumber && \vec E_j^d = \sum_b\sum_{i\neq
j=1}^{N_b}\left[3\frac{(\vec d_i\cdot\vec r_{ij})} {r_{ij}^3}\vec r_{ij} -
\frac{1} {r_{ij}^3}\vec d_{i}\right], \eea
$r_{ij}=|\vec r_{i}-\vec r_{j}|$  is a distance between ions.
Damping dispersion  functions $f_{ab}(r_{ij})$ can be determined
according to Ref.\cite{42}. It is important to notice that
interaction of polarized ions has a central-asymmetric character.
Obviously, effects related with induced rotational degrees of
freedom are expected to be small comparing to the induced dipoles
processes in the general hydrodynamics picture of ionic melts.

Nonequilibrium states of the ion-polarization model of ionic melts
are described by the nonequilibrium statistical operator
$\rho(x^N;t)$, which satisfies the Liouville equation
\bea\label{8} \frac{\partial}{\partial
t}\rho(x^N;t)+iL_N\rho(x^N;t)=0, \eea
where $iL_N$ stands for the Liouville operator, corresponding to
Hamiltonian Eq.~(\ref{2}):
\bea\label{10} iL_{N} &=& \sum_a\sum_{j=1}^{N_a}\left( \frac {\vec
p_j} {m_a}\cdot \frac{\partial} {\partial \vec r_j} +(\vec
w_j\times\hat{\vec d}_j)\cdot\frac{\partial} {\partial \hat{\vec
d}_j}\right)\nonumber
\\ &&- \sum_a\sum_{j=1}^{N_a} \left(\frac{\partial U_\mathrm{ion}} {\partial \vec r_j}
\cdot \frac{\partial} {\partial \vec p_j}+\hat{\vec d}_{j}
\times\frac{\partial U_\mathrm{ion}} {\partial \hat{\vec d}_j}
\cdot\frac{\partial} {\partial
{\stackrel{\leftrightarrow}{\mathbf{J}}}_j\cdot\vec w_j}\right),
\eea
$\hat{\vec d}_j=\frac{\vec d_j}{|\vec d_j|}$ is unit vector
describing spatial orientations of ionic dipole.

To solve the Liouville Eq.~(\ref{8}) we use the Zubarev
nonequilibrium statistical operator method, in which solutions are
searched according to the N.~Bogolyubov's idea of a reduced
description of nonequilibrium processes based on the set of
observed variables $\langle\hat P_n(\vec r)\rangle^t$. Using this
method, the solution of Eq.~(\ref{8}) can be presented in a
general form with taking into account the projection:
\bea\label{12}
\rho(x^N;t)=\rho_\mathrm{rel}(x^N;t)-\int_{-\infty}^te^{\varepsilon(t'-t)}T_\mathrm{rel}(t',t)
[1-P_\mathrm{rel}(t')]iL_N\rho_\mathrm{rel}(x^N;t')\mathrm{d}
t',\eea
where
$$
T_\mathrm{rel}(t',t)=\exp_{+}\left\{-\int_{t'}^t[1-P_\mathrm{rel}(t'')]iL_N\mathrm{d}t''\right\}
$$
is the evolution operator with taking into account Kawasaki-Gunton
projection $P_\mathrm{rel}(t)$. Projection operator depends on a
structure of relevant statistical operator
$\rho_\mathrm{rel}(x^N;t)$
\bea\label{13}
P_\mathrm{rel}(t)\rho'&=&\left(\rho_\mathrm{rel}(x^N;t)-\sum_n
\int \mathrm{d}\vec r\frac{\partial\rho_\mathrm{rel}(x^N;t)}
{\partial\langle \hat P_n(\vec r)\rangle^t}\langle \hat P_n(\vec
r)\rangle^t\right)\textrm{Tr}\{\rho'\}\nonumber \\ &&+\sum_n\int
\mathrm{d}\vec r\frac{\partial\rho_\mathrm{rel}(x^N;t)}
{\partial\langle \hat P_n(\vec r)\rangle^t}\textrm{Tr}\{\hat
P_n(\vec r)\rho'\} \eea
and possess the following properties
\begin{align*}
&P_\mathrm{rel}(t)P_\mathrm{rel}(t)=P_\mathrm{rel}(t),&
&[1-P_\mathrm{rel}(t)]P_\mathrm{rel}(t')=0,\\
&P_\mathrm{rel}(t)\rho(t)=\rho_\mathrm{rel}(t),&
&P_\mathrm{rel}(t)\rho_\mathrm{rel}(t')=\rho_\mathrm{rel}(t).
\end{align*}
Relevant statistical operator $\rho_\mathrm{rel}(t)$ is received
from the conditions of an informational entropy extremum at fixed
values of the reduced description parameters $\langle \hat
P_n(\vec r)\rangle^t$ including the normalization condition
$\textrm{Tr}\{\rho_\mathrm{rel}(x^N;t)\}=1$. Within the Gibbs approach
one can obtain
\bea\label{14}
\rho_\mathrm{rel}(x^N;t)&=&\exp\left\{-\Phi(t)-\sum_n\int
\mathrm{d}\vec r F_n(\vec r;t)\hat P_n(\vec r)\right\}, \\
\Phi(t)&=&\ln \textrm{Tr}\:\exp\left\{-\sum_n\int \mathrm{d}\vec r
F_n(\vec r;t)\hat P_n(\vec r)\right\}.\nonumber\eea
$\Phi(t)$ is the Massieu-Planck functional, $F_n(\vec r;t)$ are
the Lagrange multipliers, which are determined from the
self-consistency conditions
\bea\label{15} \langle\hat P_n(\vec r)\rangle^t = \langle\hat
P_n(\vec r)\rangle^t_\mathrm{rel} \eea
and thermodynamic relations
\bea\label{16} \frac{\delta \Phi(t)} {\delta F_n(t)} = \langle\hat
P_n(\vec r)\rangle^t, \qquad \frac{\delta S(t)} {\delta
\langle\hat P_n(\vec r)\rangle^t} = -F_n(t). \eea
Here, $S(t)$ denotes entropy of nonequilibrium state of the system
determined according to Gibbs
\bea\label{17} S(t)=\Phi(t)+\sum_n F_n(t)\langle\hat P_n(\vec
r)\rangle^t. \eea
We will consider hydrodynamic state of ionic melt within the
formulated ion-polarization model. For its description the
averaged values of densities of ionic number $\hat n^a(\vec
r)=\sum_{j=1}^{N_{a}}\delta(\vec r - {\vec r}_{j})$, their
momentum $\hat{\vec p}^a(\vec r)=\sum_{j=1}^{N_{a}}{\vec p}_{j}
\delta(\vec r - {\vec r}_{j})$, angular momentum $\hat{\vec
s}^a(\vec r)=\sum_{j=1}^{N_{a}}{\vec w}_{j}\delta(\vec r - {\vec
r}_{j})$, total energy $\hat \varepsilon(\vec
r)=\sum_a\sum_{j=1}^{N_a}\big(\frac{p_j^2}{2m_a} %
+\frac{1}{2}{\vec w}_j^\mathsf{T}{\stackrel{\leftrightarrow}{\mathbf{J}}}_j
{\vec w}_j\big)\delta(\vec r - {\vec
r}_{j})+\sum_a\sum_{j=1}^{N_a}U_\mathrm{ion}^{a}({\vec
r}_{j})\delta(\vec r - {\vec r}_{j})$ along with an induced dipole
moment $\vec d^a(\vec r)=\sum_{j=1}^{N_{a}}{\vec d}_{j}\delta(\vec
r - {\vec r}_{j})$, which are observable variables and satisfy
corresponding conservational laws, can be chosen for the reduced
description parameters. For such a set of variables
$\rho_\mathrm{rel}(x^N;t)$ can be written down as follows:
\bea\label{18} \rho_\mathrm{rel}(x^N;t) &=& \textrm{exp}
\bigg\{-\Phi(t)-\int d\vec r\beta(\vec
r;t)\Big[\hat\varepsilon(\vec r) - \sum_a\vec v^a(\vec r;t)\cdot
\hat{\vec p}^a(\vec r)\nonumber
\\ &&-\sum_a\nu^a(\vec r;t)\hat n^a(\vec r)- \sum_a\vec
E(\vec r;t)\cdot \vec d^a(\vec r) - \sum_a\vec \Omega^a(\vec
r;t)\cdot \hat{\vec s}^a(\vec r) \Big]\bigg\}, \eea
where the Lagrange multipliers $\beta(\vec r;t)$, $\vec v^a(\vec
r;t)$, $\nu^a(\vec r;t)$, $\vec E(\vec r;t)$, $\vec \Omega^a(\vec
r;t)$ are determined from the self-consistency conditions
\bea\label{19} \langle \hat\varepsilon(\vec r)\rangle^t &=&
\langle \hat\varepsilon(\vec r)\rangle^t_\mathrm{rel}, \nonumber
\\\nonumber
\langle \hat{\vec p}^a(\vec r)\rangle^t &=& \langle \hat{\vec
p}^a(\vec r)\rangle^t_\mathrm{rel}, \\\nonumber \langle \hat
n^a(\vec r)\rangle^t &=& \langle \hat n^a(\vec
r)\rangle^t_\mathrm{rel},
\\\nonumber \langle \hat{\vec d}^a(\vec r)\rangle^t &=& \langle
\hat{\vec d}^a(\vec r)\rangle^t_\mathrm{rel},
 \\ \langle \hat{\vec
s}^a(\vec r)\rangle^t &=& \langle \hat{\vec s}^a(\vec
r)\rangle^t_\mathrm{rel} \eea
and from the nonequilibrium thermodynamic relations
[Eqs.~(\ref{16}), (\ref{17})]. They have the following meaning:
$\beta(\vec r;t) =
 {1}/{k_\mathrm{B}T(\vec r;t)}$ is the inverse local temperature;
$\vec v^a(\vec r;t)$ is the mean value of hydrodynamic velocity of
ions; $\vec \Omega^a(\vec r;t)$ denotes mean value of angular
velocity of polarized ion, $\nu^a(\vec r;t) = \mu^a_{el}(\vec
r;t)+\frac{1}{2}m_a\vec v^a(\vec r;t)^2+\frac{1}{2}{\vec
\Omega}^{a\mathsf{T}}(\vec
r;t){\stackrel{\leftrightarrow}{\mathbf{J}}}{}_a{\vec
\Omega}^{a}(\vec r;t)$; $\mu^a_{el}(\vec r;t) = \mu^a(\vec
r;t)+Z_{a}e\varphi(\vec r;t)$ stands for the electrochemical
potential; $\mu^a(\vec r;t)$ is the chemical potential of  ions;
$\varphi(\vec r;t)$ denotes scalar potential of electric field
$\vec E(\vec r;t)$ induced by ions and dipoles of the system.
Electric field $\vec E(\vec r;t)=\langle\hat{\vec E}(\vec
r)\rangle^{t}$ satisfies the averaged Maxwell equations:
\bea\label{20a} \vec{\nabla}\times\langle\hat{\vec E}(\vec
r)\rangle^{t}=-\frac{1}{c}\frac{\partial }{\partial
t}\langle\hat{\vec B}(\vec r)\rangle^{t},
\eea
\bea\label{20b} \vec{\nabla}\times\langle\hat{\vec H}(\vec
r)\rangle^{t}&=&\frac{1}{c}\frac{\partial }{\partial
t}\langle\hat{\vec D}(\vec r)\rangle^{t}+
 \frac{4\pi}{c}\left(\sum_{a}\frac
{Z_{a}e}{m_{a}}\langle\hat{\vec p}^{a}(\vec r)\rangle^{t}+
 \sum_{a}\frac{1}{m_{a}}{\vec d}^{a}\cdot\vec{\nabla}\langle\hat{\vec p}^{a}(\vec
 r)\rangle^{t}\right),
 \eea
 \bea\label{20c}
\vec{\nabla}\cdot\langle\hat{\vec B}(\vec r)\rangle^{t}=0,
 \eea
\bea\label{20d} \vec{\nabla}\cdot\langle\hat{\vec D}(\vec
r)\rangle^{t}=4\pi\left(\sum_{a}Z_{a}e \langle\hat{n}^{a}(\vec
r)\rangle^{t}+ \sum_{a}{\vec d}^{a}\cdot\vec{\nabla}
\langle\hat{n}^{a}(\vec r)\rangle^{t}\right),
 \eea
Where, microscopic electric  $\hat{\vec E}(\vec r)$ and magnetic
$\hat{\vec H}(\vec r)$ fields and corresponding inductions
$\hat{\vec D}(\vec r)$, $\hat{\vec B}(\vec r)$ satisfy the
microscopic Lorenz-Maxwell equations. Known integral relations
between $\langle\hat{\vec D}(\vec r)\rangle^{t}$ and
$\langle\hat{\vec E}(\vec r)\rangle^{t}$ as well as between
$\langle\hat{\vec B}(\vec r)\rangle^{t}$ and $\langle\hat{\vec
H}(\vec r)\rangle^{t}$ determine spatially inhomogeneous
dielectric function $\epsilon ({\vec r},{\vec r}';t,t')$ and
magnetization $\chi ({\vec r},{\vec r}';t,t')$ which describe
polarization processes in the system. Acting by the operators
$[1-P_\mathrm{rel}(t)]$ and $iL_N$ on $\rho_\mathrm{rel}(t)$ in
Eq.~(\ref{12}) we obtain
\bea\label{20}
\lefteqn{(1-P_\mathrm{rel}(t))iL_N\rho_\mathrm{rel}(t)
=\bigg\{-\int \mathrm{d}\vec r\beta(\vec r;t)I_{\varepsilon}(\vec
r;t)}\nonumber\\\nonumber &&\mbox{}+\sum_a\int \mathrm{d}\vec
r\beta(\vec r;t)\vec v^a(\vec r;t) I_p^a(\vec r;t)+\sum_a\int
\mathrm{d}\vec r\nu^a(\vec r;t)I^a_n(\vec r;t)\\
&&\mbox{}+\sum_a\int \mathrm{d}\vec r\beta(\vec r;t)\vec E(\vec
r;t) I_d^a(\vec r;t) - \sum_a\int \mathrm{d}\vec r \beta(\vec r;t)
\vec \Omega^a(\vec r;t')I_s^a(\vec r;t)\bigg\}\rho_\mathrm{rel}(t),
\eea
where
\bea\label{21} &&I_{\varepsilon}(\vec r;t)=[1-P(t)]iL_N\hat
\varepsilon(\vec r),\nonumber\\\nonumber && I_{p}^a(\vec
r;t)=[1-P(t)]iL_N\hat{\vec p}^a(\vec r),\\\nonumber &&I_{n}^a(\vec
r;t)=[1-P(t)]iL_N\hat n^a(\vec r),\\\nonumber && I_{d}^a(\vec
r;t)=[1-P(t)]iL_N\hat{\vec d}^a (\vec r),\\ &&I_s^a(\vec
r;t)=[1-P(t)]iL_N\hat{\vec s}^a(\vec r), \eea
are the generalized flows, and $I^a_n(\vec r;t)=0$. $P(t)$ denotes
the generalized Mori projection operator which in this case has
the following structure
\bea\label{22}\nonumber P(t)\hat A &=&\langle \hat A
\rangle^t_\mathrm{rel} + \int \mathrm{d}\vec r \bigg\{ \frac {\delta
\langle \hat A \rangle^t_\mathrm{rel}} {\delta\langle \hat
\varepsilon(\vec r) \rangle^t}[\hat\varepsilon(\vec r)-\langle\hat
\varepsilon(\vec r) \rangle^t]\\\nonumber &&+\sum_a \frac {\delta
\langle \hat A \rangle^t_\mathrm{rel}} {\delta\langle \hat {\vec
p}^a(\vec r) \rangle^t}[\hat{\vec p}^a(\vec r)-\langle\hat{\vec
p}^a(\vec r) \rangle^t] + \sum_a \frac {\delta \langle \hat A
\rangle^t_\mathrm{rel}} {\delta\langle \hat {n}^a(\vec r)
\rangle^t}[\hat{n}^a(\vec r)-\langle\hat{n}^a(\vec r)
\rangle^t]\\
 &&+\sum_a \frac
{\delta \langle \hat A \rangle^t_\mathrm{rel}} {\delta\langle \hat
{\vec d}^a(\vec r) \rangle^t}[\hat{\vec d}^a(\vec
r)-\langle\hat{\vec d}^a(\vec r) \rangle^t] +\sum_a \frac {\delta
\langle \hat A \rangle^t_\mathrm{rel}} {\delta\langle \hat {\vec
s}^a(\vec r) \rangle^t}[\hat{\vec s}^a(\vec r)-\langle\hat{\vec
s}^a(\vec r) \rangle^t]\bigg\} \eea
and properties $P(t)[1-P(t)]=0$, $P(t)\hat P_n(\vec r)=\hat
P_n(\vec r)$.

Taking into account Eq.~(\ref{20}) we obtain the nonequilibrium
statistical operator of the ion-polarization model of ionic melt
\bea\label{23}
\nonumber\rho(t)&=&\rho_\mathrm{rel}(t)+\int_{-\infty}^t
\textrm{e}^{\varepsilon(t-t')}T_\mathrm{rel}(t,t')\int
\mathrm{d}\vec r\bigg\{\beta(\vec r,t')I_{\varepsilon}(\vec
r;t')\\\nonumber &&-\sum_a\beta(\vec r;t')\vec v^a(\vec r;t')
I^a_p(\vec r;t')-\sum_a \beta(\vec r;t')\vec E(\vec r;t')
I^a_d(\vec r;t')\\ &&-\sum_a \beta(\vec r;t')\vec \Omega^a(\vec
r;t') I^a_s(\vec r;t')\bigg\}\rho_\mathrm{rel}(t')\mathrm{d}t',
\eea
in which the generalized flows of energy density
$I_{\varepsilon}(\vec r;t)$, momentum density $ I_p^a(\vec r;t)$,
angular momentum density $ I^a_s(\vec r;t)$ and dipole moments
density $ I^a_d(\vec r;t)$ describe the dissipative processes in
the system. Using the NSO Eq.~(\ref{23}) we can obtain generalized
hydrodynamics equations for the reduced description parameters
$\langle\tilde P(\vec r)\rangle^t$ = \{$\langle \hat{n}^a(\vec
r)\rangle^t$, $\langle \hat{\vec p}^a(\vec r)\rangle^t$, $\langle
\hat{\vec s}^a(\vec r)\rangle^t$, $\langle \hat{\vec d}^a(\vec
r)\rangle^t$, $\langle \hat{\varepsilon}(\vec r)\rangle^t$\}
within the ion-polarization model of ionic melt. We present them
in matrix form
\bea\label{24} \frac{\mathrm{d}}{\mathrm{d}t}\langle \tilde P(\vec
r)\rangle^t=\langle \dot{\tilde P}(\vec
r)\rangle^t_\mathrm{rel}+\int \mathrm{d}\vec
r'\int_{-\infty}^t\textrm{e}^{\varepsilon(t-t')}\tilde\varphi_{II}(\vec
r,\vec r';t,t')\tilde F(\vec r';t')\mathrm{d}t', \eea
where $\tilde P(\vec r)$ is the column vector, $\dot{\tilde
P}(\vec r) = iL_N\tilde P(\vec r)$, $\tilde F(\vec r';t')$ =
\{$-\beta(\vec r';t')\nu^a(\vec r';t')$, $-\beta(\vec r';t')\vec
v^a(\vec r';t')$, $-\beta(\vec r';t')\vec \Omega^a(\vec r';t')$,
$-\beta(\vec r';t')\vec E(\vec r';t')$, $-\beta(\vec r';t')$\} is
the column vector of the nonequilibrium thermodynamic parameters,
$\tilde\varphi_{II}(\vec r,\vec r';t,t')$ is the matrix of the
generalized transport kernels (memory functions):
\bea\label{25} \tilde\varphi_{II}(\vec r,\vec r';t,t') = \langle
\tilde I(\vec r;t)T_\mathrm{rel}(t,t')\tilde I^{(+)}(\vec
r';t')\rangle^{t'}_\mathrm{rel}=\left(%
\begin{array}{ccccc}
  0 & 0 & 0 & 0 & 0 \\
  0 & \tilde\varphi_{pp} & \tilde\varphi_{ps} & \tilde\varphi_{pd} & \tilde\varphi_{p\varepsilon} \\
  0 & \tilde\varphi_{sp} & \tilde\varphi_{ss} & \tilde\varphi_{sd} & \tilde\varphi_{s\varepsilon} \\
  0 & \tilde\varphi_{dp} & \tilde\varphi_{ds} & \tilde\varphi_{dd} & \tilde\varphi_{d\varepsilon} \\
  0 & \tilde\varphi_{\varepsilon p} & \tilde\varphi_{\varepsilon s} &\tilde\varphi_{\varepsilon d} & \tilde\varphi_{\varepsilon\varepsilon} \\
\end{array}%
\right)_{(\vec r,\vec r';t,t')}.
 \eea
 \bea\label{28} \tilde\varphi_{pp}(\vec r,\vec r';t,t') = \left(%
\begin{array}{cc}
 \varphi^{++}_{pp} & \varphi^{+-}_{pp}\\
 \varphi^{-+}_{pp} & \varphi^{--}_{pp}\\
 \end{array}%
 \right)_{(\vec r,\vec r';t,t')}, \eea
 \bea\label{29}
 \varphi^{ab}_{pp}{(\vec r,\vec r';t,t')}=\langle I^a_p(\vec
 r;t)T_\mathrm{rel}(t,t')I^b_p(\vec r';t')\rangle^{t'}_\mathrm{rel}
  \eea
are the generalized transport kernels describing viscous
processes, herewith, $\varphi_{pp}^{++}$ and $\varphi_{pp}^{--}$
define generalized coefficients of viscosity caused by
translational motion of positively and negatively charged
polarized ions ($a,b=\{+,-\}$).
\bea\label{30} \tilde\varphi_{ss}(\vec r,\vec r';t,t') = \left(%
\begin{array}{cc}
 \varphi^{++}_{ss} & \varphi^{+-}_{ss}\\
 \varphi^{-+}_{ss} & \varphi^{--}_{ss}\\
 \end{array}%
 \right)_{(\vec r,\vec r';t,t')}, \eea
\bea\label{31} \varphi^{ab}_{ss}{(\vec r,\vec r';t,t')}=\langle
I^a_s(\vec r;t)T_\mathrm{rel}(t,t')I^b_s(\vec
r';t')\rangle^{t'}_\mathrm{rel}
 \eea
are generalized transport kernels describing viscous ionic
processes caused by rotational motion of polarized ions,
$\varphi^{++}_{ss}$ and $\varphi^{--}_{ss}$ define the generalized
coefficients of rotational viscosity of positively and negatively
charged polarized ions.
\bea\label{30*} \tilde\varphi_{dd}(\vec r,\vec r';t,t') = \left(%
\begin{array}{cc}
 \varphi^{++}_{dd} & \varphi^{+-}_{dd}\\
 \varphi^{-+}_{dd} & \varphi^{--}_{dd}\\
 \end{array}%
 \right)_{(\vec r,\vec r';t,t')}, \eea
\bea\label{31*} \varphi^{ab}_{dd}{(\vec r,\vec r';t,t')}=\langle
I^a_d(\vec r;t)T_\mathrm{rel}(t,t')I^b_d(\vec
r';t')\rangle^{t'}_\mathrm{rel},
 \eea
denote the generalized transport kernels that describe transport
processes of dipole moments of polarized ions of corresponding
species, where
\bea\label{32} I^a_d(\vec r;t)=[1-P(t)]iL_N\vec d^a(\vec r)
=[1-P(t)]\left(-\frac{\partial}{\partial\vec
r}\cdot{\stackrel{\leftrightarrow}{\mathbf{\mathbf{\Pi}}}}{}^a_d(\vec r)+ \sum_{j=1}^{N_{a}}({\vec
w}_{j}\times{\vec d}_{j})\delta(\vec r -{\vec r}_{j})\right)
 \eea
with ${\stackrel{\leftrightarrow}{\mathbf{\mathbf{\Pi}}}}{}^a_d(\vec r)$ is a tensor whose components are
$[{\stackrel{\leftrightarrow}{\mathbf{\mathbf{\Pi}}}}{}^a_d(\vec
r)]^{\alpha\beta}=\frac{1}{m_a}\sum_{j=1}^{N_a} d^\alpha_j
p_j^{\beta}\delta(\vec r-\vec r_j)$.
\bea\label{35} \tilde\varphi_{\varepsilon\varepsilon}(\vec r,\vec
r';t,t')=\langle I_{\varepsilon}(\vec
r;t)T_\mathrm{rel}(t,t')I_{\varepsilon}(\vec
r';t')\rangle^{t'}_\mathrm{rel}, \eea
is the generalized transport kernel of total energy, which
determines the generalized coefficient of heat conductivity of
ionic melt within the ion-polarization model. The matrix elements
$\tilde \varphi_{\rho p}$, $\tilde \varphi_{\rho d}$,  $\tilde
\varphi_{p d}$, $\tilde \varphi_{p \varepsilon}$, $\tilde
\varphi_{d \varepsilon}$ in~Eq.~(\ref{25}) describe dissipative
cross-correlations between momenta, dipole moments and total
energy flows of the system. The obtained nonequilibrium
statistical operator Eq.~(\ref{23}) and the generalized
hydrodynamics Eqs.~(\ref{24}) together with the set of the
generalized Maxwell Eqs.~(\ref{20a})--(\ref{20d}) are valid for
both weak and strong nonequilibrium processes in ionic melts with
taking into account polarization processes. Set of Eqs.~(\ref{24})
is unclosed and describe viscous, heat and polarization processes.
The system of hydrodynamic equations can be significantly
simplified and becomes closed for a weakly nonequilibrium
processes, when the thermodynamic parameters $\tilde F(\vec r;t)$
and the reduced description parameters $\langle \tilde P(\vec
r)\rangle^t$ slowly vary in time and space and slightly deviate
from their equilibrium values $\tilde F_0(\vec r)$, $\langle
\tilde P(\vec r)\rangle_0$ respectively. In following section we
will consider a weakly nonequilibrium processes in ionic melts
within the ion-polarization model.

\section{Generalized molecular hydrodynamics equations
for ionic melts with taking into account polarization effects}

In the case when the nonequilibrium thermodynamic parameters
$\tilde F(\vec r;t)$ slightly deviate from their equilibrium
values $\tilde F_0(\vec r)$, the relevant statistical operator
Eq.~(\ref{18}) can be expanded in deviations $\delta \tilde F(\vec
r;t) = \tilde F(\vec r;t) - \tilde F_0(\vec r)$ with taking into
account linear terms only. Then, excluding $\tilde F(\vec r;t)$
from the relevant statistical operator by means of the
self-consistency conditions [Eqs.~(\ref{19})] one can obtain
\bea\label{36} \rho_\mathrm{rel}^0(t)=\rho_0\left[1+\sum_{\vec
k}\delta\tilde P(\vec k;t)\tilde \Phi^{-1}(\vec k)\tilde P(\vec
k)\right]. \eea
Here, $\rho_0$ is the equilibrium statistical operator of ionic
melt, $\delta \tilde P(\vec k;t) = \langle\tilde P(\vec
k;t)\rangle^t - \langle\tilde P(\vec k;0)\rangle_0$, $\tilde
P(\vec k)$  is a column vector whose elements are the
Fourier-components of the reduced description parameters $\{\tilde
P(\vec k)$ = $\int\textrm{e}^{i\vec k\vec r}\tilde P(\vec r)d\vec
r\}$ = \{$ \hat{n}^a(\vec k)$,  $ \hat{\vec p}^a(\vec k)$,
$\hat{\vec s}^a(\vec k)$,$ \hat{\vec d}^a(\vec k)$, $
\hat{\varepsilon}(\vec k)$\}. $\tilde \Phi^{-1}(\vec k)$ is the
inverse of the matrix of equilibrium correlation functions $\tilde
\Phi(\vec k)$
\bea\label{37} \tilde \Phi(\vec k) = \left(%
\begin{array}{ccccc}
  \tilde\Phi_{nn}  & 0 & 0 & \tilde\Phi_{nd} & \tilde\Phi_{n\varepsilon} \\
    0  & \tilde\Phi_{pp} & 0 & 0 & 0 \\
  0  & 0 & \tilde\Phi_{ss} & 0 & 0 \\
  \tilde\Phi_{dn}  & 0 & 0 & \tilde\Phi_{dd} & \tilde\Phi_{d\varepsilon} \\
  \tilde\Phi_{\varepsilon n} & 0 & 0 & \tilde\Phi_{\varepsilon d}& \tilde\Phi_{\varepsilon\varepsilon} \\
\end{array}%
\right)_{(\vec k)}. \eea
Here,
\bea\label{38} \tilde\Phi_{nn}(\vec k) = \left(%
\begin{array}{cc}
  \Phi_{nn}^{++}(\vec k) & \Phi_{nn}^{+-}(\vec k) \\
  \Phi_{nn}^{-+}(\vec k) & \Phi_{nn}^{--}(\vec k) \\
\end{array}%
\right), \eea
is the matrix of static structure factors of ionic subsystem
$\Phi_{nn}^{ab}(\vec k)$ = $\langle \hat n^a(\vec k)\hat n^b(-\vec
k)\rangle_0$ = $S_{ab}(\vec k)$, where $\langle \ldots \rangle_0 =
\int d\Gamma_N \ldots \rho_0$.
\bea\label{41} \tilde\Phi_{pp}(\vec k) = \left(%
\begin{array}{cc}
  \Phi_{pp}^{++}(\vec k) & 0 \\
  0 & \Phi_{pp}^{--}(\vec k) \\
\end{array}%
\right), \eea
with $\Phi_{pp}^{aa}(\vec k) = \langle \hat{\vec p}^a(\vec
k)\hat{\vec p}^a(-\vec k)\rangle_0$, and
\bea\label{41*} \tilde\Phi_{ss}(\vec k) = \left(%
\begin{array}{cc}
  \Phi_{ss}^{++}(\vec k) & 0 \\
  0 & \Phi_{ss}^{--}(\vec k) \\
\end{array}%
\right), \eea
whose elements $\Phi_{ss}^{aa}(\vec k) = \langle \hat{\vec
s}^a(\vec k)\hat{\vec s}^a(-\vec k)\rangle_0$. Similarly,
\bea\label{42}\tilde\Phi_{dd}(\vec k) = \left(%
\begin{array}{cc}
  \Phi_{dd}^{++}(\vec k) & \Phi_{dd}^{+-}(\vec k) \\
  \Phi_{dd}^{-+}(\vec k) & \Phi_{dd}^{--}(\vec k) \\
\end{array}%
\right), \eea
where $\Phi_{dd}^{ab}(\vec k) = \langle \hat{\vec d}^a(\vec
k)\hat{\vec d}^b(-\vec k)\rangle_0$ are the equilibrium
correlation functions of Fourier-components of dipole moments
densities for polarized ions $a$ and $b$. 
\bea\label{44}\tilde{\Phi}_{\varepsilon\varepsilon}(\vec k) =
\langle \hat\varepsilon(\vec k) \hat\varepsilon(-\vec k)\rangle_0
\eea
is the equilibrium correlation Kubo-like function of
Fourier-components of total energy density of ionic melt. Other
matrix elements in Eq.~(\ref{37}) describe static correlations of
variables $ \hat{n}^a(\vec k)$,  $ \hat{\vec p}^a(\vec k)$, $
\hat{\vec d}^a(\vec k)$, $ \hat{\varepsilon}(\vec k)$.

In the approximation defined by Eq.~(\ref{36}) the nonequilibrium
statistical operator has the following structure
\bea\label{45} \nonumber\rho(t) &=& \rho^0_\mathrm{rel}(t)-\sum_{\vec
k}\int_{-\infty}^t\textrm{e}^{\varepsilon(t-t')}T^0_\mathrm{rel}(t,t')\\\nonumber
&&\times \bigg(I_{\varepsilon}^0(\vec k)[\tilde{\Phi}^{-1}(\vec
k)]_{\varepsilon\varepsilon}\delta\varepsilon(\vec
k;t')+\sum_{ab}\big\{I_p^a(\vec k)[\tilde\Phi^{-1}(\vec
k)]_{pp}^{ab}\delta\vec p^b(\vec k;t') \\ &&+I_d^a(\vec
k)[\tilde\Phi^{-1}(\vec k)]_{dd}^{ab}\delta\vec d^b(\vec
k;t')\big\}+\sum_{ab}I_s^a(\vec k)[\tilde\Phi^{-1}(\vec
k)]_{ss}^{ab}\delta\vec s^b(\vec k;t')\bigg)\rho_0\mathrm{d}t', \eea
where
\begin{align}\label{46} &I_{d}^a(\vec k)=(1-P_0)iL_N\hat{\vec d}^a(\vec
k),& &I_{p}^a(\vec k)=(1-P_0)iL_N\hat{\vec p}^a(\vec
k),\nonumber\\
&I_{s}^a(\vec k)=(1-P_0)iL_N\hat{\vec s}^a(\vec k), &
&I_{\varepsilon}(\vec k)=(1-P_0)iL_N\hat\varepsilon(\vec k)
\end{align}
are the generalized flows in a weakly nonequilibrium case. $P_0$
denotes Mori projection operator
\bea\label{47} P_0\hat A = \langle\hat A\rangle_0 + \sum_{\vec
k}\langle \hat A \ \tilde P^{(+)}(-\vec k)\rangle_0 \tilde
\Phi^{-1}(\vec k)\tilde P(\vec k) \eea
that possesses the following properties: $P_0(1-P_0)=0$,
$P_0\tilde P(\vec k)=\tilde P(\vec k)$. The set of the generalized
hydrodynamics equations for ionic melt within the ion-polarization
model Eq.~(\ref{45}) is now closed and can be presented in the
matrix form
\bea\label{48} \frac{\mathrm{d}}{\mathrm{d}t}\langle \tilde P(\vec k)\rangle^t -
i\tilde \Omega(\vec k)\langle \tilde P(\vec k)\rangle^t +
\int_{-\infty}^t\textrm{e}^{\varepsilon(t-t')}\tilde\varphi(\vec
k;t,t')\langle \tilde P(\vec k)\rangle^{t'}\mathrm{d}t' =0. \eea
Here,
\bea\label{49} i\tilde \Omega(\vec k) = \langle iL_N\tilde P(\vec
k)\ \tilde P^{(+)}(-\vec k)\rangle_0\tilde\Phi^{-1}(\vec k) \eea
is the frequency matrix whose elements
\bea\label{50} i\tilde
\Omega(\vec k) = \left(%
\begin{array}{ccccc}
  0  & i\tilde \Omega_{np} & i\tilde \Omega_{ns} & 0 & 0 \\
  i\tilde \Omega_{pn} & 0 & 0 & i\tilde \Omega_{pd} & i\tilde \Omega_{p\varepsilon} \\
  i\tilde \Omega_{sn} & 0 & 0 & i\tilde \Omega_{sd} & i\tilde \Omega_{s\varepsilon} \\
  0  & i\tilde \Omega_{dp} & i\tilde \Omega_{ds} & 0 & 0 \\
  0  & i\tilde \Omega_{\varepsilon p} & i\tilde \Omega_{\varepsilon s} & 0 & 0 \\
\end{array}%
\right)_{(\vec k)} \eea
are the normalized static correlation functions.
\bea\label{51} &&\tilde\varphi(\vec k;t,t') = \langle\tilde I(\vec
k)T_0(t,t')\tilde I^{(+)}(-\vec
k)\rangle_0\tilde\Phi^{-1}(\vec k) =\left(%
\begin{array}{ccccc}
  0 & 0 & 0 & 0 & 0\\
  0 & \tilde\varphi_{pp} & \tilde\varphi_{ps} & \tilde\varphi_{pd} & \tilde\varphi_{p\varepsilon} \\
  0 & \tilde\varphi_{sp} & \tilde\varphi_{ss} & \tilde\varphi_{sd} & \tilde\varphi_{s\varepsilon} \\
  0 & \tilde\varphi_{dp} & \tilde\varphi_{ds} & \tilde\varphi_{dd} & \tilde\varphi_{d\varepsilon} \\
  0 & \tilde\varphi_{\varepsilon p} & \tilde\varphi_{\varepsilon s} & \tilde\varphi_{\varepsilon d} & \tilde\varphi_{\varepsilon\varepsilon} \\
\end{array}%
\right)_{(\vec k;t,t')}
\eea
is the matrix of transport kernels (memory function), which
describe a weakly nonequilibrium transport processes including
viscous, polarization and heat processes in ionic melts within the
ion-polarization model. $\tilde I(\vec k)=\{I_{d}^a(\vec
k),I_{p}^a(\vec k),I_{s}^a(\vec k),I_{\varepsilon}(\vec k)\}$ is
the column vector, $\tilde I^{(+)}(-\vec k)=\{I_{d}^a(-\vec
k),I_{p}^a(-\vec k),I_{s}^a(-\vec k),I_{\varepsilon}(-\vec k)\}$ is
the row vector. As can be shown, in the framework of the NSO
method\cite{Zub1,Zub2}, the time correlation functions of the
basic set of the dynamic variables
\bea\label{52} \tilde\Phi(\vec k;t) = \langle \tilde P(\vec
k;t)\tilde P^{(+)}(-\vec k)\rangle_0 \eea
satisfy the Eq.~(\ref{48}) as well
\bea\label{53} \frac{\mathrm{d}}{\mathrm{d}t} \tilde \Phi(\vec k;t) - i\tilde
\Omega(\vec k) \tilde \Phi(\vec k;t) +
\int_{-\infty}^t\textrm{e}^{\varepsilon(t-t')}\tilde\varphi(\vec
k;t,t')\tilde \Phi(\vec k;t')\mathrm{d}t' =0 \eea
which takes into account memory effects. In the Markovian
approximation when these effects are negligible, in the limit
$\vec k \rightarrow 0$, $\omega \rightarrow 0$, the set of
Eqs.~(\ref{53}) can be presented as follows:
 \bea\label{54}
\frac{\mathrm{d}}{\mathrm{d}t}\tilde\Phi(\vec k;t) + \tilde T(\vec k)\tilde\Phi(\vec
k;t) = 0, \eea
where
\bea\label{55} \tilde T(\vec k) = -i\tilde \Omega(\vec k)
+\tilde\varphi(\vec k) \eea
is the generalized hydrodynamic matrix. The system of
Eqs.~(\ref{53}) permits to study the time correlation functions of
partial dynamics in ionic melts. Though if the elements of the
frequency matrix in Eq.~(\ref{50}) can be calculated via
equilibrium characteristics, in particular, the partial structure
factors, the calculation of the transport kernels Eq.~(\ref{51})
is a well-known problem and it can be implemented only
approximately with taking into account special feature of
dissipative processes in the system. This issue accompanies the
investigation of collective excitation spectrum for ionic melts
with taking into account polarization effects, in particular in
the hydrodynamics limit $\vec k\rightarrow 0,\omega\rightarrow 0$.
In the following section we use obtained above the generalized
molecular hydrodynamics equations for ionic melts, in particular,
within a viscoelastic approximation, to investigate the spectrum
of collective excitations in the hydrodynamic limit.

\section{Collective modes of ionic melts. Viscoelastic approximation.}

In order to study collective excitations within the
ion-polarization model of molten salts we will use viscoelastic
approximation and will consider a longitudinal modes only. We will
not take into account total energy density as well as rotational
degrees of freedom of induced dipoles and pass from the
description in terms of partial dynamic variables
$\{\hat{n}^a(\vec k),\ \hat{\vec p}^a(\vec k), \ \hat{\vec
d}^a(\vec k)\}$ to the description based on total densities of
ions mass $\hat{n}(\vec k)=m_{+}\hat{n}^{+}(\vec
k)+m_{-}\hat{n}^{-}(\vec k)$, total ions momentum $\hat{\vec
p}(\vec k)=\hat{\vec p}^{+}(\vec k)+\hat{\vec p}^{-}(\vec k)$,
total charge $\hat{\varrho}(\vec k)=Z_{+}e\hat{n}^{+}(\vec
k)+Z_{-}e\hat{n}^{-}(\vec k)$ and density of total current of
charge $\hat{\vec J}(\vec k)=({Z_{+}e}/{m_{+}})\hat{\vec
p}^{+}(\vec k)+({Z_{-}e}/{m_{-}})\hat{\vec p}^{-}(\vec k)$. In
addition we consider only negatively charged atoms to be
polarized, i.e. negative ions possess induced dipole moments.
Thus, we include density of dipole moment of negatively charged
ions $\hat{\vec d}^{-}(\vec k)$ to the set of the reduced
description parameters. However, it is more convenient to use the
variable $\hat{\vec d}(\vec k)$ orthogonal to the variables
$\hat{n}(\vec k)$ and $\hat{\varrho}(\vec k)$:
\bea \label{var-d} \hat{\vec d}(\vec k)=\hat{\vec d}^{-}(\vec k)
-\vec A(\vec k)\hat{n}(\vec k)-\vec B(\vec k)\hat{\varrho}(\vec
k). \eea
Here, $A(\vec k)$ and $B(\vec k)$ are determined by the relations
\bea {\vec A}(\vec k)=\langle\hat{\vec d}^-(\vec k)\hat{n}(-\vec
k)\rangle_0[\tilde\Phi^{-1}(\vec k)]_{nn}+\langle\hat{\vec
d}^-(\vec k)\hat{\varrho}(-\vec k)\rangle_0[\tilde\Phi^{-1}(\vec
k)]_{\varrho n}\,, \eea
\bea  {\vec B}(\vec k)=\langle\hat{\vec d}^-(\vec k)\hat{n}(-\vec
k)\rangle_0[\tilde\Phi^{-1}(\vec k)]_{n\varrho}+\langle\hat{\vec
d}^-(\vec k)\hat{\varrho}(-\vec k)\rangle_0[\tilde\Phi^{-1}(\vec
k)]_{\varrho \varrho}\,. \eea
In this case, the generalized hydrodynamic matrix has the
following form
\bea \label{577} T(\vec k)=\left(
\begin{array}{c c c c c}
0 & -i\Omega_{np} & 0 & -i\Omega_{nJ} & 0\\
-i \Omega_{pn} &  \varphi_{pp} & -i\Omega_{p\varrho} &
-i\Omega_{p\varrho}+\varphi_{p\varrho}
& -i\Omega_{pd}+\varphi_{pd}\\
0 & 0 & 0 & -i\Omega_{\varrho J} & 0 \\
 -i \Omega_{Jn} & -i\Omega_{Jp}+\varphi_{Jp} & -i \Omega_{J\varrho} &  \varphi_{JJ}
 &  -i\Omega_{Jd}+\varphi_{Jd} \\
0 &  -i\Omega_{dp}+\varphi_{dp} & 0 & -i\Omega_{dJ}+\varphi_{dJ}
 &  \varphi_{dd}\\
\end{array} \right)_{(\vec k)}
\eea
with the corresponding elements of the frequency matrix and the
matrix of memory functions Eq.~(\ref{51}) constructed on the
generalized flows $\tilde{I}_{\cal A}(\vec k)=(1-P_0)iL_N
\tilde{\cal A}(\vec k)$ in Markovian approximation. Here, $P_0$ is
t he Mori-like projection operator Eq.~(\ref{47}) constructed on
the set of dynamic variables
\begin{eqnarray}\label{577.2}
\tilde{\cal A}(\vec k)=\{\hat{n}(\vec k), \hat{\vec p}(\vec k),
\hat{\varrho}(\vec k), \hat{\vec J}(\vec k), \hat{\vec d}(\vec
k)\}.
\end{eqnarray}
The wave vector $\vec{k}=(0,0,k)$ ($k=|\vec k|$) is directed along
the $OZ$ axis. $\varphi_{pp}(k)=k^2\phi_{pp}=k^2\eta/mn$, where
$\eta$ is a viscosity coefficient of molten salt. Total current of
charge is not conserved quantity and we can write down $iL_N\vec
J(\vec k)={\vec J}_0+i\vec k\cdot {\stackrel{\leftrightarrow}{\mathbf{\mathbf{\Pi}}}}_J(\vec k)$.
Therefore, in the $k \rightarrow 0$ limit we have
$\varphi_{JJ}(\vec k)=\phi_{JJ}+O(k^2)$, where
$\phi_{JJ}={\omega_{p}^{2}}/{4\pi \sigma}$ with $\sigma$ is an
electroconductivity coefficient of molten salt.
$\omega_{p}^{2}=\sum_{a}\omega_{a}^{2}$, where
$\omega_{a}^{2}=4\pi {n_{a}(Z_{a}e)^{2}}/{m_{a}}$ is a squared
plasma frequency of ions of species $a$. Similar is a situation with
the total dipole moment: $iL_N\vec d^-(\vec k)=\vec w^-(\vec
k)+{i\vec k}\cdot{\stackrel{\leftrightarrow}{\mathbf{\mathbf{\Pi}}}}{}_d^-(\vec k)$. Here, $\vec w^-(\vec
k)=\sum_{l=1}^{N_-}\vec w_l \times \vec d_l e^{i\vec k {\vec
r}_l}$ and ${\stackrel{\leftrightarrow}{\mathbf{\mathbf{\Pi}}}}{}_d^-(\vec k)=\sum_{l=1}^{N_-}\vec p_l \vec
d_l e^{i\vec k {\vec r}_l}/m_{-}$ are the Fourier transforms of
quantities defined in Eq.~(\ref{32}). Therefore, we obtain that
the transport kernel $\varphi_{dd}(\vec k)=\varphi_{d^- d^-}(\vec
k)=D_0+O(k^2)$, where $D_0=\lim_{\vec k \to
0}\int_0^{\infty}\langle\vec w^-(\vec k;t)\cdot\vec w^-(-\vec
k)\rangle_0[{\tilde\Phi}^{-1}(\vec
k)]_{dd}\mathrm{d}t$ is the normalized
coefficient of rotational diffusion of polarized ions.

The elements $i\Omega_{\cal A \cal A}(\vec k)$ in the hydrodynamic
matrix Eq.~(\ref{577}) can easily be calculated applying the $\vec k
\rightarrow 0$ limit. Taking into account $\langle \hat{\vec
p}^{+} \cdot\hat{\vec p}^{+}\rangle_{0}=\lim_{\vec k\rightarrow
0}\langle \hat{\vec p}^{+}(\vec k)\cdot \hat{\vec p}^{+}(-\vec
k)\rangle_{0}={n_{+}m_{+}}/{\beta}$, $\langle \hat{\vec p}^{-}
\cdot\hat{\vec p}^{-}\rangle_{0}=\lim_{\vec k\rightarrow 0}\langle
\hat{\vec p}^{-}(\vec k)\cdot \hat{\vec p}^{-}(-\vec
k)\rangle_{0}={n_{-}m_{-}}/{\beta}$ ($n_{+}={N_{+}}/{V}$,
$n_{-}={N_{-}}/{V}$ are the equilibrium average value of
concentrations of negative and positive ions), we find for
$\langle \hat{\vec p} \cdot\hat{\vec p}\rangle_{0}$ the following
result $\lim_{\vec k\rightarrow 0}\langle \hat{\vec p}(\vec k)\cdot
\hat{\vec p}(-\vec
k)\rangle_{0}=({n_{+}m_{+}}/{\beta}+{n_{-}m_{-}}/{\beta})$,
or $\Phi_{pp}=\langle \hat{\vec p} \cdot\hat{\vec
p}\rangle_{0}={m}/{\beta}$, $m=n_{+}m_{+}+n_{-}m_{-}$ is the
equilibrium value of the total mass density of ions of the system.
Similarly, one can find $\Phi_{jj}=\langle \hat{\vec J}\cdot
\hat{\vec J}\rangle_{0}=\lim_{\vec k\rightarrow 0}\langle \hat{\vec
J}(\vec k) \cdot\hat{\vec J}(-\vec
k)\rangle_{0}={\omega_{p}^{2}}/{4\pi \beta}$.
Cross-correlation function is equal to $\Phi_{pj}=\langle
\hat{\vec p} \cdot\hat{\vec J}\rangle_{0}=\lim_{\vec k\rightarrow
0}\langle \hat{\vec p}(\vec k)\cdot \hat{\vec J}(-\vec
k)\rangle_{0}=(Z_{+}en_{+}+Z_{-}en_{-})/{\beta}$, and due
to the electroneutrality condition $(Z_{+}en_{+}+Z_{-}en_{-})=0$
one can obtain $\Phi_{pj}=\Phi_{jp}=0$. Considering the structure
of the static correlation functions and values of the averages
$\langle \hat{\vec p}^{+} \cdot\hat{\vec p}^{+}\rangle_{0}$,
$\langle \hat{\vec p}^{-} \cdot\hat{\vec p}^{-}\rangle_{0}$, we
find for $i\Omega_{\cal A\cal A}(\vec k)$ the following results:
\bea\label{691} i\Omega_{np}=ik, \qquad
i\Omega_{pn}=ik\frac{m}{\beta}[\tilde{\Phi}^{-1}]_{nn}=ik c_T^2,
\eea
where $c_T$ is the isothermal sound velocity,
\bea\label{692} i\Omega_{nJ}=0, \qquad
i\Omega_{Jn}=ik\frac{\omega_{p}^2}{4\pi\beta}[\tilde{\Phi}^{-1}]_{\varrho
n}=ik\omega_{Jn}, \eea
\bea\label{693 } i\Omega_{\varrho p}=0, \qquad
i\Omega_{p \varrho}
=ik\frac{m}{\beta}[\tilde{\Phi}^{-1}]_{n\varrho}=ik\omega_{p
\varrho}, \eea
\bea\label{694} i\Omega_{\varrho J}=ik, \qquad
i\Omega_{J \varrho }=ik\frac{\omega_{p}^{2}}{4\pi
\beta}[\tilde{\Phi}^{-1}]_{\varrho \varrho}=ik^{-1}\Omega_q^2\,
.\eea
In the expressions Eqs.~(\ref{691})--(\ref{694})
$i\Omega_{nJ}=i\Omega_{\varrho p}=0$ due to condition of
electroneutrality of the system whereas $\omega_{Jn}$,
$\omega_{p\varrho}$ and $\Omega_{q}^2$ tend to nonzero constant in
the $\vec k\rightarrow 0$ limit. The quantities
$[{\tilde\Phi}^{-1}]_{nn}$, $[{\tilde\Phi}^{-1}]_{n\varrho}$,
$[{\tilde\Phi}^{-1}]_{\varrho n}$, $[{\tilde\Phi}^{-1}]_{\varrho
\varrho}$ as well as $[{\tilde\Phi}^{-1}]_{dd}$ are the elements
of the matrix ${\tilde\Phi}^{-1}$ inverse to the matrix of static
correlation functions
\bea \label{578} {\Phi}=\left(
\begin{array}{c c c c c}
\Phi_{nn} & 0 & \Phi_{n \varrho} & 0 & 0\\
0 & \Phi_{pp} & 0 & 0 & 0\\
\Phi_{\varrho n} & 0& \Phi_{\varrho \varrho} & 0 & 0 \\
0 & 0 & 0 & \Phi_{JJ}  & 0 \\
0 & 0 & 0 & 0 & \Phi_{dd}\\
\end{array} \right),
\eea
with $\Phi_{dd}=\lim_{\vec k\rightarrow 0}\langle \hat{\vec d}(\vec k)
\cdot\hat{\vec d}(-\vec k)\rangle_{0}$.
Whereas, $\Phi_{nn}=\lim_{\vec k\rightarrow 0}\Phi_{nn}(\vec
k)=\lim_{\vec k\rightarrow 0}\{m_{+}^2S^{++}(\vec
k)+m_{+}m_{-}[S^{+-}(\vec k)+S^{-+}(\vec k)]+m_{-}^2S^{--}(\vec
k)\}=m_{+}^2S^{++}(0)+m_{+}m_{-}[S^{+-}(0)+S^{-+}(0)]+m_{-}^2S^{--}(0)$,
where we use $S^{++}(\vec k)$, $S^{+-}(\vec k)$, $S^{-+}(\vec k)$,
$S^{--}(\vec k)$ for the partial static structure factors. In
particular, $S^{++}(0)=1+n_{+}\int \mathrm{d}
\vec{r}(g_{2}^{++}(\vec{r})-1)=\chi^{+}_{T}{n_{+}}/{\beta}$,
$S^{--}(0)=1+n_{-}\int \mathrm{d} \vec{r}(g_{2}^{--}(\vec{r})-1)=
\chi^{-}_{T}{n_{-}}/{\beta}$, where $g_{2}^{++}(\vec{r})$,
$g_{2}^{++}(\vec{r})$, $\chi^{+}_{T}$, $\chi^{-}_{T}$ are the
equilibrium pair distribution functions of ions and their
isothermal susceptibilities. Similarly, one can calculate
``charge-density'' $\Phi_{n \varrho}$, $\Phi_{\varrho n}$ and
``charge-charge'' $\Phi_{\varrho \varrho}$ correlation functions.
We find $\Phi_{n \varrho}=Z_{+}e[m_{+}S^{++}(0)+m_{-}S^{+-}(0)]+
Z_{-}e[m_{+}S^{-+}(0)+m_{-}S^{--}(0)]$, $\Phi_{\varrho
\varrho}=(Z_{+}e)^{2}S^{++}(0)+ Z_{+}Z_{-}e^{2}[S^{+-}(0)+
S^{-+}(0)]+(Z_{-}e)^{2}S^{--}(0)$. In the $\vec k\rightarrow 0$ limit,
$\Phi_{nn}(\vec k)$ tends to nonvanishing constant, while $\Phi_{n\varrho}(\vec k)$
and $\Phi_{\varrho\varrho}(\vec k)$ have a $k^2$ asymptote\cite{12}.

Then, the same way one can obtain the elements $i\Omega_{pd}$,
$i\Omega_{dp}$, $i\Omega_{dj}$, $i\Omega_{jd}$ of the frequency
matrix
\bea\label{697}
i\Omega_{pd}=ik\left(\frac{n_{-}}{\beta}\vec{d}^{-} -\vec
A(0)\frac{m}{\beta}\right)[{\tilde\Phi}^{-1}]_{dd}=ik\omega_{pd}\,,
\eea
\bea\label{697.2}
i\Omega_{dp}=ik\left(\frac{n_{-}}{\beta}\vec{d}^{-}-\vec
A(0)\frac{m}{\beta}\right)
[{\tilde\Phi}^{-1}]_{pp}=ik\omega_{dp}\,, \eea
\bea\label{698}
i\Omega_{Jd}=ik\left(\frac{Z_{-}en_{-}}{\beta}\vec{d}^{-} -\vec
B(0)\frac{\omega_p^2}{4\pi\beta}\right)[{\tilde\Phi}^{-1}]_{dd}=ik\omega_{Jd}\,,
\eea
\bea\label{698.2}
i\Omega_{dJ}=ik\left(\frac{Z_{-}en_{-}}{\beta}\vec{d}^{-} -\vec
B(0)\frac{\omega_p^2}{4\pi\beta}\right)[{\tilde\Phi}^{-1}]_{JJ}=ik\omega_{dJ}\,.
\eea

In order to obtain the collective excitation spectrum of ionic
melt with taking into account polarization properties we have to
solve the equation
\bea \label{699} \rm{det}\left(
\begin{array}{c c c c c}
z(k) & -i k & 0 & 0 & 0\\
-i kc_T^2 &  z(k)+k^2\phi_{pp} & -ik\omega_{p\varrho} & i k D_{pJ}
& -i k D_{pd}\\
0 & 0 & z(k) & -i k & 0 \\
 -ik \omega_{Jn} & i k D_{Jp} & -i k^{-1}\Omega_{q}^2 &  z(k) + \phi_{JJ}
 &  D_{Jd} \\
0 &  -i k D_{dp} & 0 & D_{dJ}
 & z(k)+ D_0\\
\end{array} \right)=0
\eea
We do not provide here expressions for the coefficients $D_{pJ}$,
$D_{Jp}$ describing cross-correlations between total current and
total current of charge since they do not enter the hydrodynamic
expressions for collective excitations, what we will see below.
``Total current~- dipole moment'' cross-correlations are described
by coefficients
\begin{align} \label{699.2}
&D_{dp}=-i\omega_{dp}+\lim_{\vec k \to 0}\int_0^{\infty}\langle\vec{w}^-(k;t)(1-P_0){\stackrel{\leftrightarrow}
{\mathbf{\mathbf{\Pi}}}}{}_{p}(-k)\rangle_0\Phi_{pp}^{-1}(\vec k)\mathrm{d}t, \\
&D_{pd}=-i\omega_{pd}-\lim_{\vec k \to
0}\int_0^{\infty}\langle(1-P_0){\stackrel{\leftrightarrow}
{\mathbf{\mathbf{\Pi}}}}{}_{p}(k)e^{-iL_Nt}\vec{w}^-(-k)\rangle_0\Phi_{dd}^{-1}(\vec k)\mathrm{d}t.
\end{align}
Here, ${\stackrel{\leftrightarrow}{\mathbf{\mathbf{\Pi}}}}{}_{p}(k)$ denotes Fourier transform of the stress
tensor introduced as  $iL_N\vec p(\vec k)=i\vec k\cdot{\stackrel{\leftrightarrow}{\mathbf{\mathbf{\Pi}}}}{}_{p}(\vec k)$. The cross-coefficients describing
correlations between charge and polarization processes, in their
turn, are defined as follows:
\begin{align} \label{699.3}
&D_{Jd}=\lim_{\vec k \to 0}\int_0^{\infty}\langle\vec{J}_0(\vec
k;t)\cdot\vec
w^-(\vec k)\rangle_0\Phi_{dd}^{-1}(\vec k)\mathrm{d}t, \\
&D_{dJ}=\lim_{\vec k \to 0}\int_0^{\infty}\langle \vec w^-(\vec
k;t)\cdot\vec{J}_0(\vec k)\rangle_0\Phi_{JJ}^{-1}(\vec k)\mathrm{d}t.
\end{align}
Coefficients $D_{Jd}$, $D_{dJ}$, $D_{pd}$, $D_{dp}$ tend to
nonzero constants in the $\vec k\to 0$ limit.

Looking for the solutions as series in wave vector $z(k)= z_0 +
z_1 k + z_2 k^2$, one can get the set of equations for evaluation
of the coefficients $z_0$, $z_1$ and $z_2$. After calculations we
find that collective excitations spectrum contains a pair of
propagating and three relaxation modes. The propagating modes are
the acoustic ones
\bea \label{701} z_{s}^{\pm}=\pm ic_Tk-\Gamma_s k^2%
\eea
with the sound velocity $c_T$ and the attenuation coefficient $\Gamma_s$
\bea \label{702}
\Gamma_s=\frac{1}{2}\left(\phi_{pp}+\frac{D_{dp}D_{pd}}{D_0}
\right). \eea
The hydrodynamic expressions for the acoustic (low-frequency)
excitations in molten salts do not contain contributions from
charge variables. It was shown\cite{43} within the GCM approach
using various sets of dynamic variables that, in the domain of
small wavenumbers, coupling between mass and charge (or
concentration) fluctuations is negligible and our result is in
agreement with it. As it follows from
Eqs.~(\ref{701})--(\ref{702}) the existence of dipole moment in
iodine ions do not effect the magnitude of sound velocity. Its in
a consistency with the results obtained by means of the concept of
``bare'' modes with the help of AI (which take into account
polarization effects) and rigid-ion MD simulations for molten NaCl\cite{24} and NaI\cite{26}. This approach yields very close to
each other ``bare'' \textsl{ab initio}  and ``bare'' rigid-ion
linear dispersions of low-frequency excitations. It indicates that
effect of polarization processes over sound velocity in negligible
what is in agreement with Eq.~(\ref{701}), however, the sound
damping coefficient is renormalized due to these effects. Neglecting the polarization
effects we simply obtain for the sound damping coefficient $\Gamma_s=\phi_{pp}/{2}$
which corresponds to a viscoelastic approximation.

Among three relaxation modes one is due to rotational dipolar
diffusion of polarized ions, shifted by $\delta$ from its
hydrodynamic value because the interaction with other relaxation
modes
\bea \label{703} z_D=-(D_0+\delta)+O(k^2). \eea

A pair of remaining relaxation modes are due to electric
conductivity $\sigma$ and are defined as follows:
\bea \label{704} z_{q}^{\pm}=-\Gamma_q \pm \Delta_q+O(k^2), \eea
where
\bea \label{705}
\Gamma_q=\frac{1}{2}\left(\phi_{JJ}-\delta\right), \qquad
\Delta_q=\sqrt{\Gamma_q^2-\left[\Omega_q^2-D_{dJ}D_{Jd}+\delta\left(D_0-2\Gamma_q\right)\right]}.
\eea
A relaxation mode due to electric conductivity in NaI was
previously obtained\cite{23} within the GCM approach with the
help of \textsl{ab initio} MD simulations, however, based on the
extended set of dynamic variables. Neglecting the contributions
from polarization processes ($D_{dJ}$, $D_{dJ}$ as well as
$\delta$ tend to zero) in Eqs.~(\ref{704})--(\ref{705}) lead us to
simple expression
\bea \label{706} z_{0,q}^{\pm}=-\frac{1}{2}\phi_{JJ} \left[1 \pm
\sqrt{1-\left({2\Omega_q}/{\phi_{JJ}}\right)^2}\right]+O(k^2) .
\eea

It also should be mentioned that calculation of eigenvectors of
the generalized hydrodynamic matrix Eq.~(\ref{699}) permits to
determine the whole set of time correlation functions
\bea \label{707} F_{\cal AA}(k;t)=\sum_{\alpha}G_{\cal
AA}^{\alpha}(k)e^{-z_\alpha(k)t} \eea
(where amplitudes $G_{\cal AA}^{\alpha}(k)$ are determined via the
eigenvectors related to eigenvalues $z_\alpha(k)$) with partial
contributions of all collective excitations and to study the
effect of polarization processes over them.

\section{Conclusions}

In summary, we presented the statistical description of hydrodynamic processes
of ionic melts with taking into account polarization effects
caused by deformation of outer electron shells of ions. It is
implemented by means of the Zubarev nonequilibrium statistical
operator method that enables to study both weak and strong
nonequilibrium processes. As the result, the nonequilibrium
statistical operator and the generalized hydrodynamics equations
with taking into account polarization effects are received within
the ion-polarization model of molten salts, when the observable
values such as the nonequilibrium averaged values of densities of
ionic number $\hat n^a(\vec r)$, their momentum $\hat{\vec
p}^a(\vec r)$, angular momentum $\hat{\vec s}^a(\vec r)$, total
energy $\hat \varepsilon(\vec r)$ along with dipole moment $\vec
d^a(\vec r)$ are chosen for the reduced description parameters.
The generalized molecular hydrodynamics equations for ionic melts
with taking into account polarization effects in weakly
nonequilibrium case are obtained as well. Based on them within the
viscoelastic approximation for molten salts in the limit
$\vec k\rightarrow 0$, $\omega\rightarrow 0$ we found the spectrum of
collective excitations consists of two conjugated acoustic and
three relaxation modes: one is due to rotational dipolar diffusion
of negatively charged ions and two relaxation modes due to
electroconductivity. Analytical results presented here are in a
qualitative agreement with the results of GCM approach based on
\textsl{ab initio} (considering polarization effects) and
rigid-ion MD simulations. Thus, taking into account polarizability of ions
do not effect dispersion of acoustic excitations and provide a shift of relaxation modes.

The model can easily be extended to the thermoviscoelastic one by
taking into account total energy density fluctuations. Herewith,
the spectrum will obtain one more hydrodynamic mode with $k^2$
asymptote due to thermal diffusivity. The coupling of viscous and
thermal processes will lead to the renormalization of sound
velocity from isothermal to adiabatic one. The respective damping
coefficient will be determined by viscothermal effects as well as
charge density fluctuations.

Knowing the spectrum of collective excitations permits to study
the time correlation functions built on the basic dynamic
variables Eq.~(\ref{577.2}) and investigate contributions caused
by polarization effects in them. Investigations of transport processes in
multi-component melts LiF, BeF$_{2}$, ThF$_{4}$, UF$_{4}$ for nuclear
reactors with the uranium cycle\cite{37c} and the melts LiF, NaF, BeF$_{2}$, ZrF$_{4}$, Li$_{2}$BeF$_{4}$ for the uranium and plutonium fuels\cite{37k} are of great interest within the presented approach. Whereas the negatively charged ions only can be considered as polarized for the simple molten alkali halides, the polarizability of all ions should be taken into account for more complex melts mentioned above.

\section*{Acknowledgements}

The authors want to thank T. Bryk for important critical comments.
Valuable comments of anonymous referees are also greatly acknowledged.

\end{document}